# New Detection System for Heavy Element Research

**Yu. S. Tsyganov** [1], A.N.Polyakov, A. A. Voinov, M.V.Shumeyko

[1] *FLNR, JINR*

**Abstract:**

New detection system design for heavy element research with $^{48}$Ca projectile has been reported. This system is based on application of 32 position sensitive strip PIPS detector and low pressure pentane filled TOF detector application in $^{48}$Ca induced nuclear reactions. To suppress beam associated background products new version of real-time method of "active correlations" has been applied. Examples of applications in $^{249}$Bk+$^{48}$Ca and $^{243}$Am+$^{48}$Ca reactions are presented. The system development to operate together (in parallel) with the digital ORNL detection system to provide a quick search for ER-alpha correlation chains has been discussed too.

## 1. Introduction

The Dubna Gas Filled Recoil Separator (DGFRS) is now mostly advanced facility in the field of synthesis of superheavy elements. It was put into operation at U400 main FLNR cyclotron in 1989. Since that time more than 50 new superheavy nuclides were synthesized at the DGFRS. From the viewpoint of the focal plane detectors application at the DGFRS many types of silicon detectors were applied since 1989. In the Table 1 these detector types are shown in chronological order ( Si(Li) 1 cm$^2$, 5mm depth in the short experiment $^{232}$Th+$^{11}$B in 1992).

**Table 1. Main detector types of the DGFRS from 1988**

| Type | Size | ~year | manufacturer | Additional info |
|---|---|---|---|---|
| n-Si(Au) | 55 mm diam | 1988-1989 | FLNR, JINR | Adjustment of the DGFRS |
| P-Si(Al) | 48 mm | 1989-1992 | FLNR, JINR | To measure Efficiency (add chamber) |
| n-Si(Au) | 18mm x 6 det array | 1990 | FLNR, JINR | $^{40}$Ar+$^{235}$U experiment (neg res) |
| n-Si (Au) | 2x2.5 cm x 6 det | 1991-1993 | FLNR, JINR | Ch. St. systematic, Some HI exp-s e.g. U+Ne, O. $^{207}$Pb+$^{40}$Ar→$^{244}$Fm+3n |

| PIPS pos. sens. | 1x4 x 12 strip | 1994-2008 | CANBERRA NV, Belgium | $^{48}$Ca beam+AcTag |
| PIPS pos. sens. | 0.37 x 6 x 32 strip | 2009 – present time | CANBERRA NV, Belgium | $^{48}$Ca+$^{249}$Bk→117 |
| DSSSD | 6x12 cm$^2$ 48x128 strips (~330 μm depth) | 2013 November, Is planned | Micron Semiconductor, UK | $^{48}$Ca +$^{240}$Pu |
| Solid state (plastic) | 14x6 cm$^2$ | 1989 | JINR | $^{207}$Pb+$^{40}$Ar→$^{244}$Fm+3n |

## 2. DGFRS detection system

From the viewpoint of detection system the experiment on synthesis and study of the properties of superheavy nuclei is one of the most difficult tasks. In fact, these experiments can be considered extreme in many cases:

- extremely low formation cross sections of the products under investigation,
- extremely high heavy ion beam intensities,
- high radioactivity of actinide targets, which are used in the experiments aimed at the synthesis of superheavy nuclei,
- extremely long duration of the experiment,
- extremely low yield of the products under investigation,
- very high required sensitivity of the detection system and
- radical suppression of the background products ( method of "active correlations"),
- high reliability level of monitoring system,
- high quality visualization system for spectroscopy on-line data flow.

Four last points are the subject of the present paper in a wide sense as well as the examples of applications of the mentioned systems and methods. Deeply modified method of "active correlation" for both PIPS and DSSSD detector is one of them.

## 2. New the DGFRS detection system (PIPS detector, CANBERRA NV. Scenario A)

New DGFRS detecting module consists of 32 strip (Fig.1) position sensitive PIPS detector (size 12 x 6 cm$^2$), 24 strip side PIPS detector and 8 strip "VETO" PIPS detector. The EVRs recoiling from the target were separated in flight from $^{48}$Ca beam ions, scattered particles, and transfer reaction products by the DGFRS. EVRs passed through a time-of-flight (TOF) detector and were implanted in focal plane PIPS detector. Namely, detecting TOF signal provides discrimination of charge particles coming from cyclotron with respect to decay signals of implanted nuclei. The position-averaged detection efficiency for α-decay of implanted nuclei was about 87% of 4π.

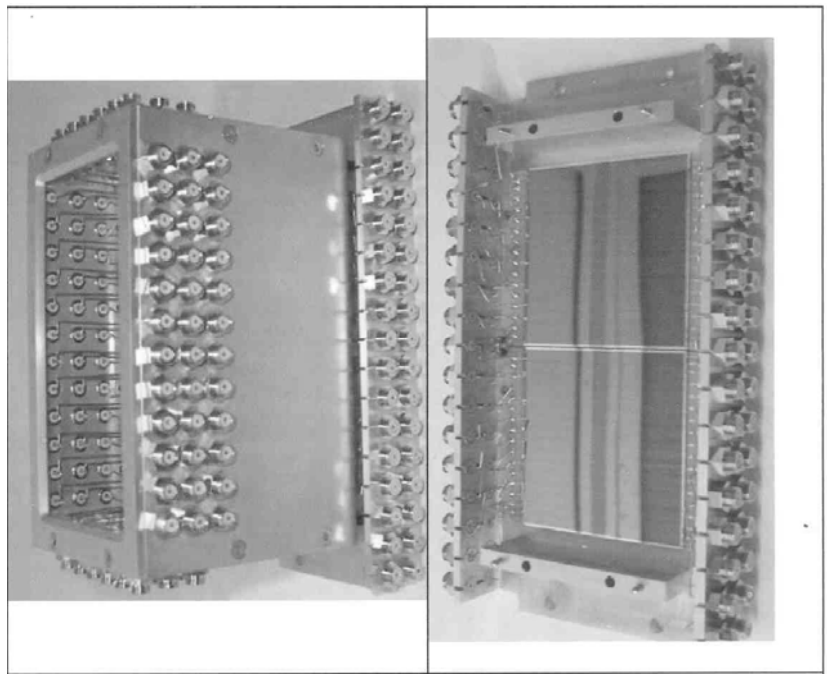

**Fig.1** Focal plane 32-strip PIPS detector the DGFRS

To measure a vertical position we use both two signals – "top" signal and "bottom" signal. For operation with α-particle range signals (~0.6 to 20 MeV) we apply fast ADC PA3n ( design by "Tekhinves", special economy zone "Dubna"). It allow to detect event by event signals with minimum (firs) dead time of 4.9 μs (see Fig.2) anr regular dead time from 12 to 40 μs. It means that "minimum" signal sequence which can be detected by the system is approximately this: 5-40-5-40-5-40 or 5-12-5-12-5-12… microseconds. The algorithm to search for EVR-alpha correlation sequence in a real-time mode is based on a simple idea. It consists in searching the time-energy –position EVR (evaporation residue) – alpha links using the discrete representation of the resistive layer of PIPS focal plane detector separately for "EVR" , "alpha " signals. Thus, the real PIPS detector is represented in the PC RAM in the form of four matrices: two for recoils(EVR, static top/bottom), another two for alphas (top/bottom, dynamic-no storage elements except for the given cells). Those matrices are filled with the values of elapsed times obtained from CAMAC "Tekhinvest" module. The second index of the matrix element is defined by event vertical position, whereas the first index is in fact the strip number (1 to 32). In a case of alpha signal detection, a comparison with recoil matrix is made, involving neighbor elements (+/- 4). If the minimum "top" or "bottom" is less or equal to the setting time (or calculated from any formulae with incoming energy dependence) the system turns on the beam chopper which deflects the heavy ion beam in the injection line of the cyclotron for one minute. At the next step PC code (Builder C++) ignores the vertical position of the alpha signal during beam-off interval. If such a decay takes place in the same strip that generated the pause, the duration of the beam-off interval is prolonged up to ten minutes (experiment $^{48}$Ca+$^{249}$Bk→117+3,4n; 2012).

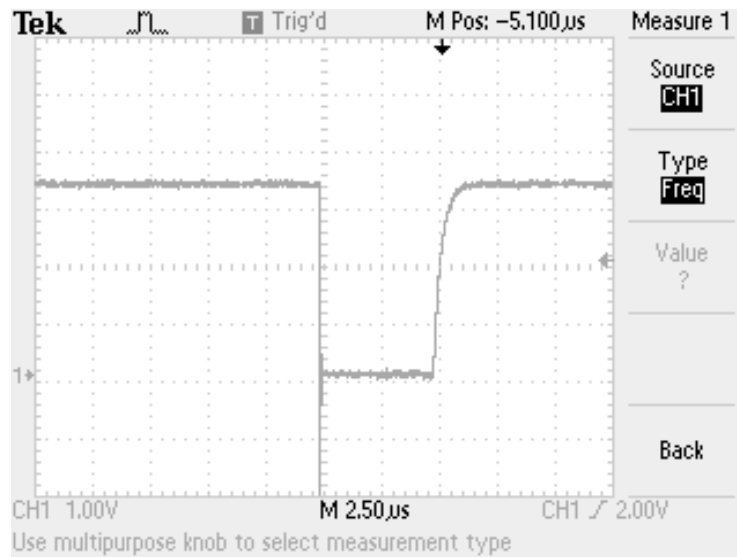

**Fig.2a** Signal "dead time" of ADC PA3n. (in principle, low level ~2.2 μs except for 4.9 in the presented case)

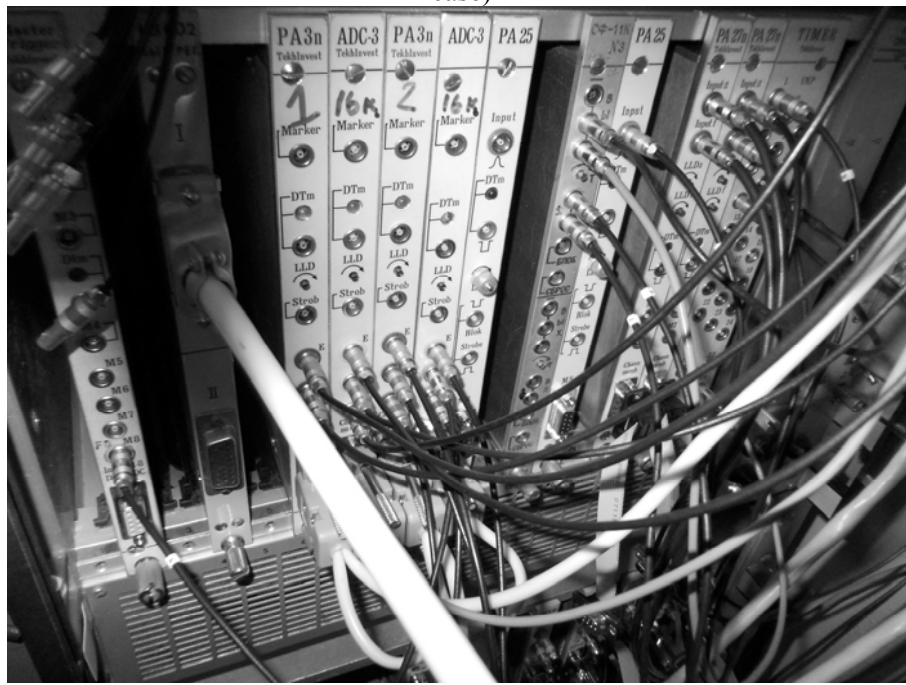

**Fig.2b** CAMAC crate containing two Pa3n ADC units.

Of course, in the real experiment, in order to apply the mentioned above algorithm, calibrations procedures are strongly required. Calibration parameters are extracted from test reaction $^{nat}Yt + {}^{48}Ca \rightarrow {}^{217}Th + 3n$. It takes about 200 calibration parameters to provide the a real-time search for EVR- α chains. Note, that the DGFRS detection system operates together with PC-based parameter monitoring and protection system which is described in detail in the Ref.[4-8]. In the Fig.3 one can see main user interface form.

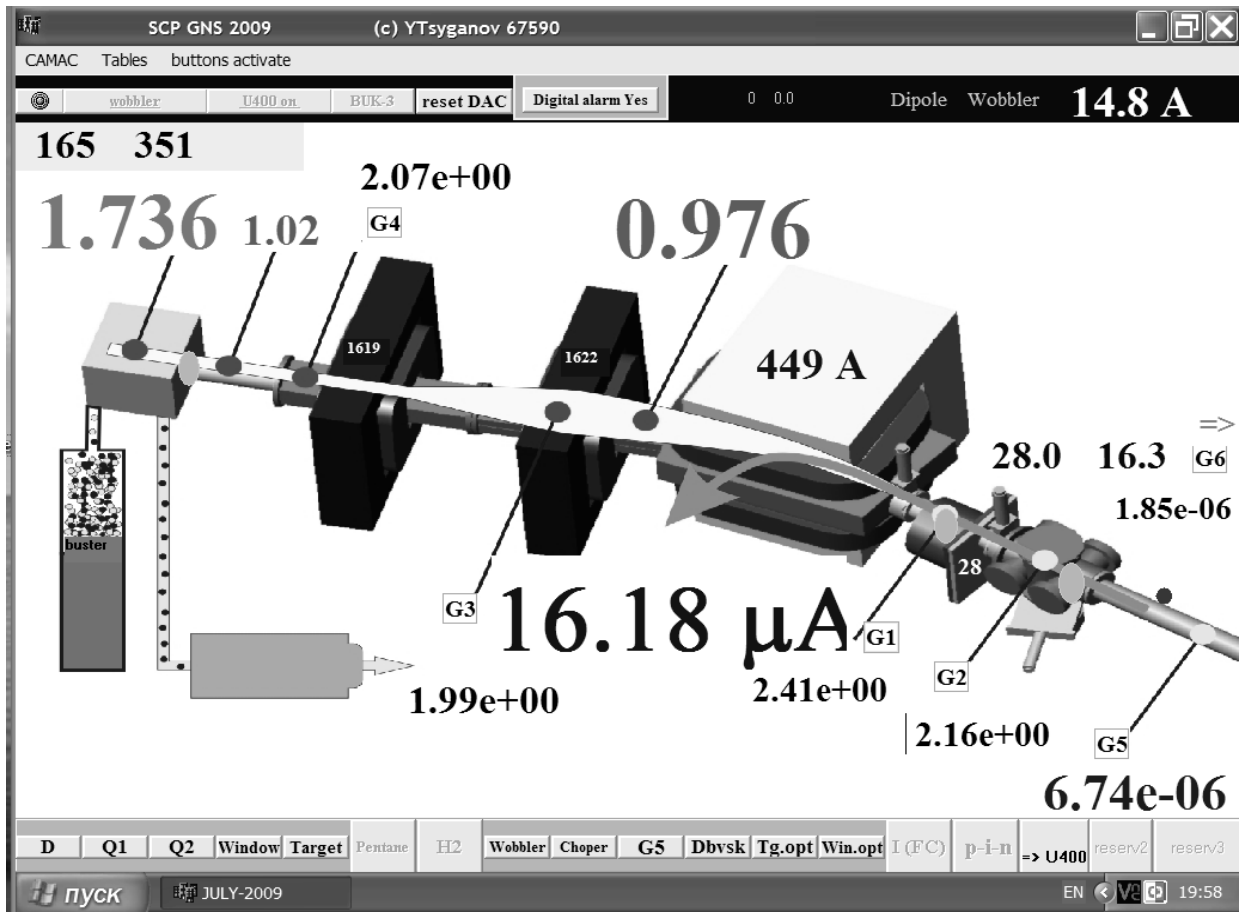

**Fig.3** The protection system user interface main form

In the above shown Fig.3 the main parameters are: 1.736- pentane pressure in the TOF module (Torr), 16.18 μA- beam current at the DGFRS Faraday cup, 28.0- target wheel rotation speed (1/s), 0.976- hydrogen pressure in the separator, 449 A –dipole magnet current value, 165 – rate of events (1/s), 6.74e-06 – vacuum value before the separator, 14.8 A – beam wobbling winding current value. The main goal of that system application is to provide quick deflection of the cyclotron beam in the case of the system detected any "alarm" situation (e.g. associated with rotation of the actinide target wheel, vakuum line, hydrogen or pentane pressure, etc.)

**2.1. Examples of application**

Examples of application of the reported system in the Bk+Ca → 117 +3n experiment, one can see in the Fig.4 multi-decay chain of Z=117 nucleus. Beam Off phase is shown by a shadow.

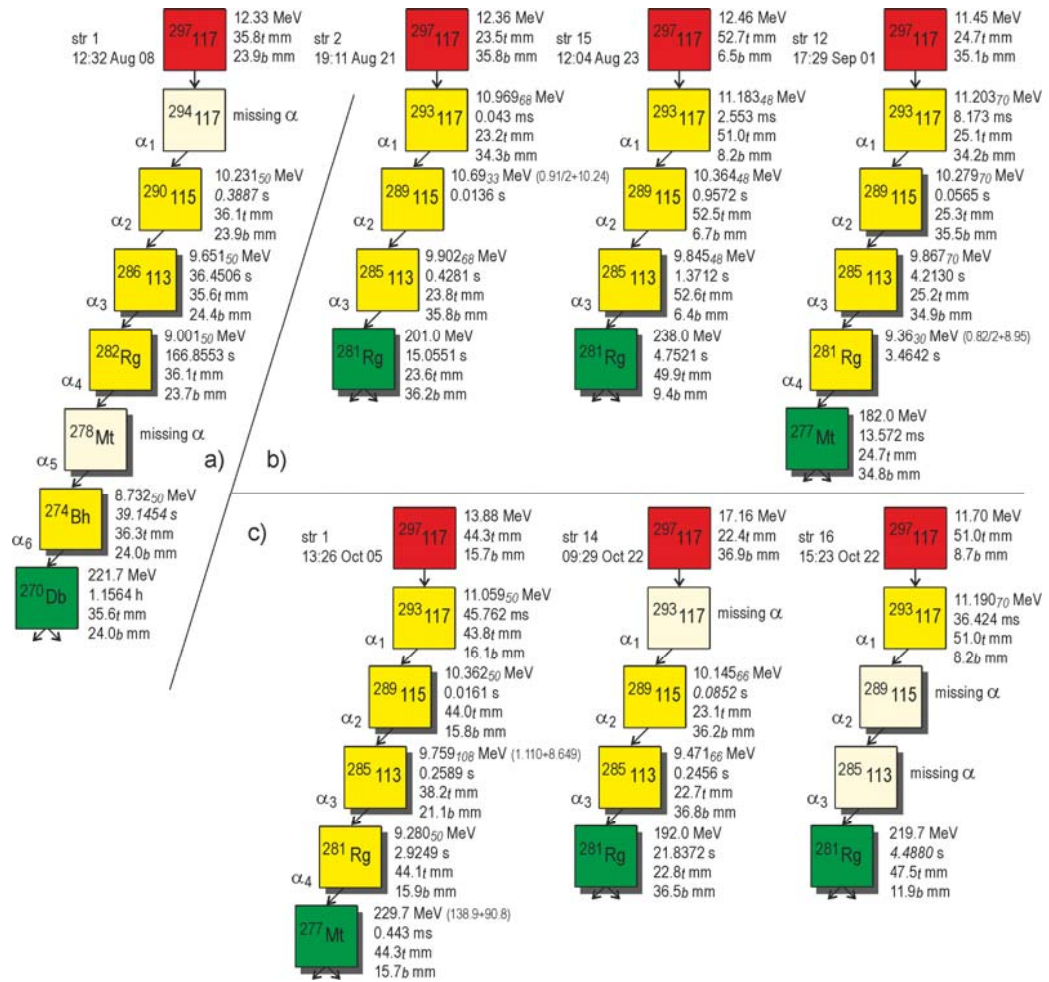

**Fig.4** Examples of Z=117 decay chains detected with the presented system.

3. **New the DGFRS detection system (DSSSD detector, Micron Semiconductors. Scenario B )**

In nearest future application in the long term experiments we plan to apply 48x128 strip DSSSD detector produced by Micron Semiconductors (UK). Now, that system is in a stage of α-particle tests. We plan to put the system into operation in the end of 2013. For data taking and searchinf for ER-alpha correlated sequences in a real-time mode, **RED_STORM** Builder 6 C++ code has been designed [7]. For data visualization (~100 histograms) Delphi 7 based **VIEV_spectr** code has been designed.

Below, in the Fig.7a main crate of the detection system for DSSSD scenario is shown, whereas in the Fig.7b first results of α-particle tests are shown for strips № 17-28.

It can be easily seen broken contact for strip №32 is observed during the test.

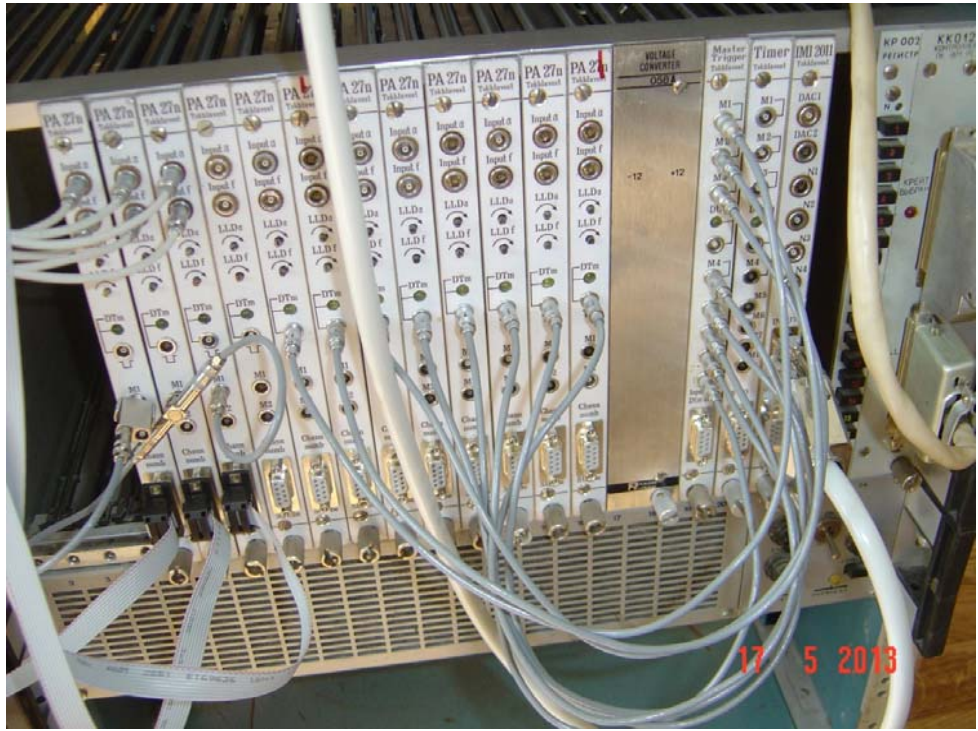

**Fig.7 a**  CAMAC crate with ADC,s for DGFRS detection system to operate in parallel with ORNL (TN, USA) digital system [9].

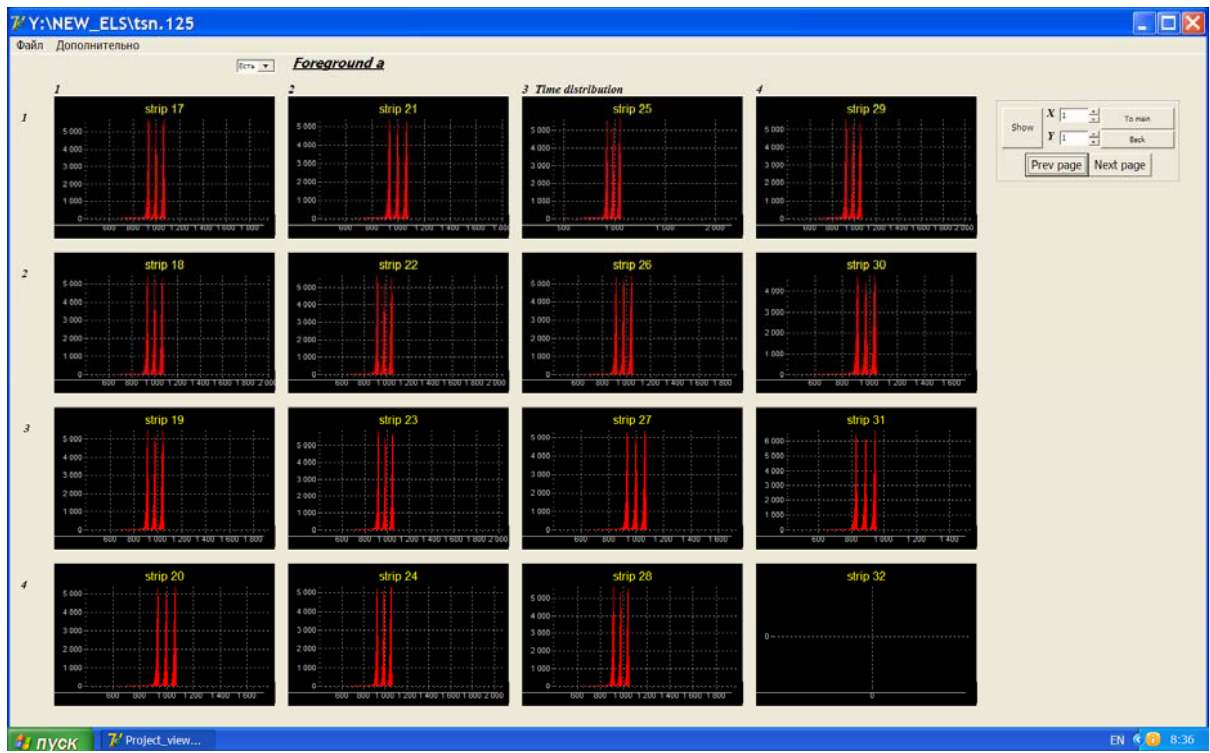

**Fig.7b**  Test spectra ($^{233}$U, $^{238}$Pu, $^{239}$Pu) for DSSSD based detection system of the DGFRS.

## 4. SUMMARY

New DGFRS detection and parameter monitoring system has been designed and successfully applied in the $^{249}Bk+^{48}Ca \rightarrow 117+3,4n$. Namely with radical background suppression due to that application, it has became possible to provide clear identification of decaying products. In a nearest future we plan to parallelize that system with ORNL design digital system [9].

Authors are indebted to Drs. V.Zhuchko and A.Sukhov for their help in putting the system into operation. This work was supported in part by RFBR Grant № 13-02-12052.